\documentclass[journal,twoside,web]{IEEEtran}
\usepackage[superscript]{cite}

\usepackage{amsmath,amssymb,amsfonts}
\usepackage{algorithmic}
\usepackage{graphicx}
\usepackage{textcomp}

\usepackage{multirow}
\usepackage{subcaption}
\usepackage{hyperref}
\usepackage{url}
\usepackage{dsfont}
\usepackage{bbold}
\usepackage{nicematrix}
\usepackage{arydshln}
\usepackage{colortbl}
\usepackage{tabulary}
\usepackage{hhline}
\usepackage{relsize}
\usepackage{balance}
\usepackage{graphicx}
\usepackage{ifpdf}
\usepackage[binary-units,per-mode=symbol]{siunitx}
\usepackage{glossaries}

\usepackage{booktabs}

\makeatletter
\renewcommand\@biblabel[1]{#1.}
\makeatother

\newacronym{PD}{PD}{Parkinson's Disease}
\newacronym{HC}{HC}{Healthy Control}
\newacronym{SHAP}{SHAP}{SHapley Additive exPlanations}
\newacronym{CV}{CV}{Cross-Validation}
\newacronym{XGBoost}{XGBoost}{eXtreme Gradient Boosting}
\newacronym{SVM}{SVM}{Support Vector Machine}
\newacronym{KNN}{KNN}{K-Nearest Neighbor}
\newacronym{RFECV}{RFECV}{Recursive Feature Elimination Cross-Validation}
\newacronym{GWS}{GWS}{Group-Wise Scaling}

\def\BibTeX{{\rm B\kern-.05em{\sc i\kern-.025em b}\kern-.08em
    T\kern-.1667em\lower.7ex\hbox{E}\kern-.125emX}}
\begin{document}
\title{Continuous-Speech Parkinson’s Disease Detection Using Acoustic and Inharmonicity Features}
\author{Rujia Li, Niloofar~Momeni, 
        Susanna Whitling, 
        Andreas Jakobsson$^{*}$
\thanks{This work was supported by the Swedish SRA ELLIIT, project C07, and was partially supported by the Wallenberg AI, Autonomous Systems and Software Program (WASP) funded by the Knut and Alice Wallenberg Foundation. The computations were enabled by resources provided by the National Academic Infrastructure for Supercomputing in Sweden (NAISS), partially funded by the Swedish Research Council through grant agreement no. 2022-06725.}%
\thanks{R.~Li, N.~Momeni, and A.~Jakobsson are with the Centre for Mathematical Sciences, Mathematical Statistics, Lund University, Box 118, SE-221 00 Lund, Sweden (e-mails: lirujiaa@gmail.com, niloofar.momeni@matstat.lu.se, andreas.jakobsson@matstat.lu.se). $^{*}$Corresponding author. }%
\thanks{S.~Whitling is with the Department of Logopedics, Phoniatrics, and Audiology, Faculty of Medicine, Lund University, Sweden (e-mail: susanna.whitling@med.lu.se).}%
}

\maketitle

\begin{abstract}
Notable efforts have been made to identify Parkinson's disease (PD) from vocal data, primarily using sustained vowel phonations. In this work, we extend on these efforts introducing a PD identification approach for continuous speech, enabling a practical background monitoring of voice data to detect vocal changes indicative of PD. 
Using two distinct data sets, we compare the best sustained vowel model with that of the proposed continuous speech model, clearly illustrating the preferential performance of the latter. We examine approaches for speaker level evaluation and data leakage preventions, as well as how vowel information may be reliably extracted from continuous speech. The proposed method framework exploits both traditional acoustic representations and a promising novel inharmonicity based framework, showing how the latter provides complementary information improving the performance for one of the data sets; however, for the other data set, this information did not significantly improve (nor reduce) the performance, suggesting that further studies are required before being able to draw firm conclusions in its use. Overall, the work clearly illustrates the benefit of forming PD classification using continuous speech compared to using sustained vowel sounds. 
\end{abstract}


\section{Introduction}
\label{sec:introduction}

Parkinson's disease (PD) is a serious health problem that currently affects millions of people around the world and continuous to grow due to the aging population.\cite{momeni2024mobile,who2023parkinson,parkinsonFoundationStats,schrag2000quality} 
The disease causes a wide range of motor and non-motor symptoms, notably including soft voice (hypophonia), lack of voice (aphonia), monotone speech, voice tremors, altered or irregular speech rate, as well as stuttering or hesitations, all of which affect the volume, intonation, articulation, and fluency of the voice and speech.\cite{brabenec2017speech,atalar2023hypokinetic,cao2025speech,kalia2015parkinsons,murman2012early} As an early detection of PD, as well as individual monitoring of the disease progression, may improve managing of the disease and slow down its progression, the topic of identifying PD from voice recordings has attracted significant attention in the recent literature. \cite{mei2021machine,moro2021advances,gullapalli2022early,momeni2024mobile,momeni2025gws,idrisoglu2026multiclass,farajollahi2026diagnosis}
Speech-based PD identification is both promising and offers the benefits of being non-invasive, repeatable, and relatively inexpensive, and offers the potential of detecting the disease before patients report speech impairment as a primary clinical complaint.\cite{hlavnicka2017connected}

So far, most of the literature has focused on PD identification using sustained vowels as the main speech task.\cite{vaiciukynas2017speech,ozbolt2022methodological,cao2025speech} This choice is understandable: sustained phonation is easy to elicit, relatively simple to standardise, and provides a stable acoustic signal from which recording-level features may be extracted directly. However, a sustained vowel mainly reflects a controlled phonatory condition. It does not fully represent the articulatory transitions, temporal organisation, and coordination demands that occur in connected or continuous speech. Studies using connected, running, or continuous
speech suggest that less constrained speech material may reveal additional PD-related information
\cite{hlavnicka2017connected,vasquezcorrea2015continuous,
orozcoarroyave2016running,appakaya2023connected,farago2023cnn,postma2025evaluating}, but such material is also more difficult to represent reliably.
In this study, we focus on vowel-centred regions as a controlled subset of continuous speech. Although the full continuous recording will contain potentially informative cues from vowels, consonants, transitions, and pauses, the vowel-dominant voiced regions generally provide locally more stable and approximately periodic signal segments, which are particularly suitable for estimating fundamental frequency, harmonic structure, and related acoustic measures. Previous studies have also shown that vowel articulation extracted from connected-speech tasks can capture Parkinson-related articulatory changes
\cite{Rusz2013Vowel,Illner2023Vowel}.
%
%

Worth noting is also that speech-based PD classification is sensitive to the evaluation design. If recordings from the same speaker are split across training and test sets, models may exploit speaker-specific characteristics rather than disease-related structure. Reported performance may also be affected by demographic imbalance, shortcut learning, and preprocessing choices that are not fitted strictly within the training data.\cite{ozbolt2022methodological,ge2023evaluation,momeni2025gws} These issues are particularly important in small clinical speech datasets, where the number of recordings can be much larger than the number of independent speakers.

This paper investigates whether continuous speech can provide additional predictive and {\em interpretable} value over a strong sustained-vowel reference when both conditions are evaluated under the same speaker-level protocol. The study uses the publicly available NeuroVoz corpus, a Castilian-Spanish PD speech dataset containing both sustained-vowel recordings and short \emph{listen-repeat} utterances from the same speaker cohort,\cite{mendes2024neurovozpaper} as well as 
the linguistically distinct Swedish Voice Diagnostics (VD) dataset, which similarly contains both sustained-phonation and continuous-speech recordings. Within each data set, this allows
the performance of controlled sustained phonation and continuous speech to be compared under the same evaluation framework, whereas comparing the results from data sets differing in both language and recording mode shows the consistency of the proposed approach across heterogeneous data conditions.

The proposed continuous-speech framework does not treat each full recording as one homogeneous signal. Instead, it first reduces the utterances to reliable vowel-centred regions. These regions are then used to form two complementary representations. The first is a recording-level acoustic representation that combines
extraction-derived summaries of the retained vowel-centred regions with short-time descriptors from the extended Geneva Minimalistic Acoustic Parameter Set (eGeMAPS), extracted using openSMILE \cite{eyben2010opensmile,eyben2016gemaps}.

The second is an inharmonicity-based representation. In voiced speech, spectral components are approximately organised around integer multiples of a fundamental frequency, although real signals need not be perfectly harmonic. Such almost-harmonic behaviour can be described through the deviations of observed spectral components from an ideal harmonic grid. 
In this work, we make use of earlier work that has developed fundamental-frequency definitions for almost-harmonic signals
\cite{elvander2020f0,elvander2023otp}, and  examine whether these harmonic deviations provide complementary information for PD classification. Here, the eGeMAPS features are modelled at the recording level and are then aggregated to the speaker level, whereas the inharmonicity features are pooled directly into a person-level representation.

The main contributions of this paper are as follows:
\begin{enumerate}
    \item We introduce an interpretable continuous speech PD identification model and compare the performance of using vowel-centred continuous-speech instead of sustained-vowels for PD detection, clearly showing the notable improved performance of the former.
    \item We introduce a novel person-level inharmonicity representation to capture complementary harmonic-offset structures in the continuous-speech material.
    \item The proposed framework is evaluated on two linguistically and task-distinct data sets. In both cases, the continuous-speech acoustic representation outperforms the corresponding sustained-vowel reference, whereas the added value of the inharmonicity representation is found to be data-set dependent.

    \item The acoustic, inharmonicity, and score-level fusion models are evaluated under a leakage-aware protocol with speaker-wise splitting, train-only preprocessing, and subject-level performance assessment.

\end{enumerate}
The remainder of the paper is organized as follows: in the next section, we introduce the data sets and the evaluation protocol. Section~\ref{sec:continuous_framework} introduces the proposed continuous speech framework, including the vowel-centred frame extraction, the novel inharmonicity features, and the resulting person-level representation. Section~\ref{sec:classification_fusion} examines the benchmarking, as well as the classification and fusion approach,  Section~\ref{sec:results} presents our results, 
and Section~\ref{sec:discussion_conclusion} our discussion of these.
Finally, Section~\ref{subsec:discussion_conclusion_final} contains our conclusions.

\begin{table}[t]
\centering
\caption{The considered NeuroVoz data set.}
\label{tab:dataset_summary}
\footnotesize
\begin{tabular}{lccc}
\toprule
 & HC & PD & Total \\
\midrule
Speakers & 55 & 52 & 107 \\
Sustained-vowel recordings & 476 & 559 & 1035 \\
Listen-repeat recordings & 867 & 828 & 1695 \\
Analysed recordings & 1343 & 1387 & 2730 \\
\midrule
Age, mean & 64.0  & 71.5  & 67.7  \\
Female recordings & 635 & 537 & 1172 \\
Male recordings & 682 & 850 & 1532 \\
\bottomrule
\end{tabular}

\vspace{0.5ex}
\raggedright
\end{table}

\section{Data and Evaluation Protocol}
\label{sec:data_protocol}

In the following classification, we will examine two distinct datasets, 
both discussed briefly in the following. This is followed by a discussion on speaker-level evaluation and bias prevention.

\subsection*{The NeuroVoz Corpus}
\label{subsec:corpus_tasks}

The NeuroVoz corpus introduced by Mendes-Laureano {\em et al.}\cite{mendes2024neurovozpaper} is a Castilian-Spanish speech dataset\footnote{Inclusion criteria are given in the paper.} containing recordings from speakers with PD as well as healthy controls (HCs). Recordings were acquired in a quiet, although non-acoustically isolated, room during the clinical routine using an
AKG C420 headset microphone connected through a phantom-powered preamplifier to a SoundBlaster Live sound card at a sampling rate of 44.1~kHz. The recording system had a reported mean signal-to-noise ratio of 24.3~dB \cite{mendes2024neurovozpaper}.  The data set consists of 2,903 recordings from 108 speakers\footnote{One PD speaker was excluded from the comparison because only spontaneous-speech recordings were available for that subject. Two HC speakers had missing age information, and one HC speaker missing sex information. These speakers were retained in the experiments as age and sex were not used as model inputs, but have been omitted in Table~\ref{tab:dataset_summary}.
}, covering sustained vowels, listen-repeat utterances, diadochokinetic tests (rapid syllable repetition), and spontaneous speech. In this work, we only examine the sustained-vowel and listen-repeat recordings. 
The sustained-vowel recordings contain the five Spanish vowels $[a]$, $[e]$, $[i]$, $[o]$, and $[u]$, whereas the listen-repeat recordings contain 16 short utterances repeated by each speaker. 
The resulting subset consists of 2,730 recordings from 107 speakers: 55 HC and 52 PD speakers. The breakdown of the subset is summarized in 
Table~\ref{tab:dataset_summary}. It is worth noting that the NeuroVoz metadata also include broader clinical
information such as disease duration, Unified Parkinson's Disease Rating Scale (UPDRS) score, Hoehn--Yahr stage, and medication information, but do not provide a standardized measure specifically quantifying dysarthria severity \cite{mendes2024neurovozpaper}.

\subsection*{The Voice Diagnostic Dataset}
\label{subsec:swedish_dataset}

The Voice Diagnostics (VD) dataset\footnote{Regrettably, the VD dataset is not publicly available due to privacy agreements and institutional data-sharing restrictions.} was developed as
part of our ongoing collaboration with Swedish clinical researchers
and physicians. 
The data set consists of sustained phonation of $[a]$ and spontaneous continuous speech recordings, from 44 PD and 136 HC speakers, with 5,959 HC recordings and 1,140 PD recordings. 
For inclusion, patients had to be at least 18 years old, have a confirmed PD diagnosis, as well as be able to make recordings using a smartphone without assistance. 
Recordings were made via a mobile app using the users' own mobile phones, which varied in brand and model, at a sampling rate of 44.1~kHz. Device information is available for the individual recordings, but  information on the recording location, environmental conditions, and quantitative background-noise level is not. The recordings were downsampled to 16~kHz prior to processing. The mean age is 68 for HC and 66 for PD subjects.
The data set is summarized in 
Table~\ref{tab:swedish_dataset_summary}.
The VD data set differs from NeuroVoz in two ways: firstly, the continuous-speech material is spontaneous rather than elicited by a fixed \emph{listen-repeat} protocol. Secondly, the number of recordings per speaker is highly uneven, ranging from 1 to 981 recordings per subject. In the following, we introduce a weighting to allow for this discrepancy. Corresponding measures of dysarthria severity and broader clinical disease severity are not available for the VD data used in this study.

\begin{table}[t]
\centering
\caption{The Voice Diagnostics (VD) data set.}
\label{tab:swedish_dataset_summary}
\footnotesize
\begin{tabular}{lccc}
\toprule
 & HC & PD & Total \\
\midrule
Speakers & 136 & 44 & 180 \\
All recordings & 5959 & 1140 & 7099 \\
Spontaneous-speech recordings & 2957 & 618 & 3575 \\
Sustained $[a]$ recordings & 3002 & 522 & 3524 \\
\midrule
Age, mean & 68 & 66 & 67.5 \\
Female recordings & 3354 & 714 & 4068 \\
Male recordings & 2605 & 426 & 3031 \\
\bottomrule
\end{tabular}
\end{table}

\subsection*{Speaker-Level Evaluation and Leakage Prevention}
\label{subsec:evaluation_protocol}

In this work, we examine speaker level PD classification in order to allow the decision to be supported by multiple recordings. To facilitate this, the evaluation used five stratified folds over speakers, and all recordings from a given speaker were assigned to the same fold.
In each fold, the validation set was formed from approximately 20\% of the training speakers, again using speaker-level splitting to ensure that no speakers occur in both training and validation sets.

This validation subset was used for model selection, early stopping where applicable, as well as fusion-weight and operating-threshold selections (as further discussed in the following). The hold-out fold was not used during any of these steps. Final recording-level predictions were aggregated to speaker level before computing the main performance measures.
The same separation principle was applied beyond the data split itself. Several preprocessing and modelling choices estimate quantities from the observed data, and these quantities can leak information from held-out speakers if they are fitted before the fold separation is applied. For this reason, feature filtering, imputation, standardisation, group-wise scaling, model tuning, fusion-weight selection, and threshold selection were all fitted within the relevant training or validation partition of each outer fold. The fitted parameters were then applied unchanged to the held-out test speakers. This fold-local design was used to avoid optimistic performance estimates caused by leakage\footnote{Here, leakage refers to the unintended use of information from held-out data during model development or preprocessing; of particular concern is speaker leakage, where
recordings from the same individual occur in both training and held-out partitions.} in the validation procedure\cite{ozbolt2022methodological,ge2023evaluation}.

Proceeding, we note that the classification results may be  confounded with non-disease related biases in the data sets, most notably by age and gender. As shown in Tables~\ref{tab:dataset_summary} and \ref{tab:swedish_dataset_summary}, the PD speakers are on average  older than the HC speakers; similarly, the gender distribution is not even.
As both age and sex notably affect the voice characteristics, a classifier trained on globally standardised features may partially learn demographic structure rather than disease-related speech patterns. To reduce this shortcut risk, group-wise scaling was applied using age and sex only to define normalisation groups, not as model inputs.\cite{momeni2025gws} Within each fold, speakers were assigned to groups defined by sex and broad age bucket\footnote{The noted HCs missing the required information were assigned to a missing-metadata group.}, and robust scaling parameters were estimated separately within each group from the training data. Medians and interquartile ranges were used rather than means and standard deviations to make the scaling less sensitive to outlying recordings. If a subgroup contained too few training observations for a stable estimate, the corresponding fold-level training statistics were used as a fallback. 
Furthermore, for the NeuroVoz continuous-speech data, the 16 fixed listen-repeat utterance tasks were standardised separately within each outer fold. For each task, the mean and standard deviation of each acoustic feature were estimated using only the training recordings, and the same parameters were then applied unchanged to the
corresponding validation and test recordings. For the VD data, the continuous-speech material consisted of a single spontaneous-speech task, so task-wise standardisation was not applicable. Instead, feature-wise z-score standardisation was performed within each outer fold, with the mean and standard deviation estimated from the training recordings only and then applied unchanged to the validation and test recordings.\cite{pedregosa2011scikit}

Finally, training samples were weighted to limit the influence of speakers with many recordings: If a speaker $s$ contributed $n_s$ training recordings, each of that speaker's recordings received a weight proportional to $1/n_s$, followed by normalisation to unit mean within the training partition. This kept the optimisation closer to the speaker-level target used for final evaluation.

\section{The Proposed Continuous-Speech Approach}
\label{sec:continuous_framework}

In this work, we introduce a continuous-speech PD identification approach. Compared to the sustained vowels, the continuous speech is  more heterogeneous and  contains more articulatory movement, timing variation, and coordination demands. The spoken sentences contain vowel nuclei, consonantal intervals, transitions, pauses, and segments with weak or unstable periodicity. If treated as one segment, the characteristics of these regions would be mixed into one representation and would therefore weaken the link between the extracted features and the voiced material that is most relevant for the present analysis.
As a result, the proposed approach first reduces each utterance to selected vowel-centred regions. These regions served two purposes. They defined the parts of the openSMILE descriptor stream used for the continuous-speech acoustic representation, and provided the short-time frames from which harmonic-offset information should be estimated. The acoustic representation is done at a  recording-level, whereas the inharmonicity representation was pooled at the person level to obtain a more stable description of the harmonic-offset variations.

\begin{figure}[t]
    \centering
    \includegraphics[width=\linewidth]{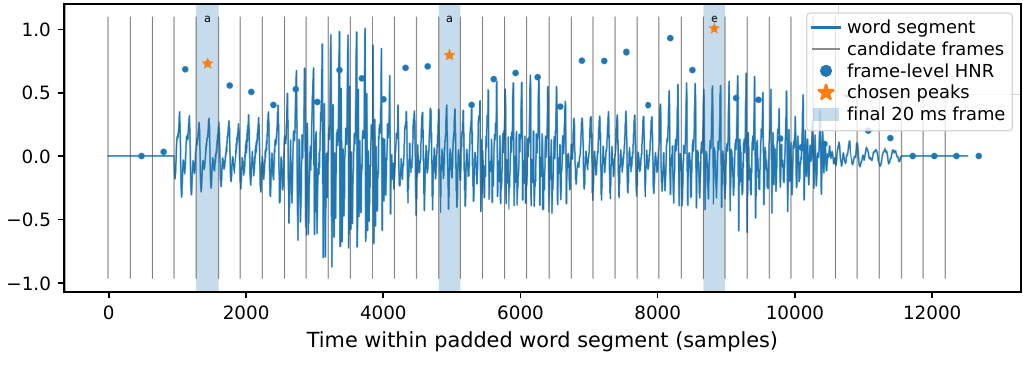}
    \caption{Example of HNR-based centre selection within a recognised
    word segment. The waveform and frame-level HNR values are normalised
    only for visualisation. Grey vertical lines indicate candidate analysis
    windows, orange markers show the local HNR peaks selected as vowel
    centres, and the shaded regions indicate the final 20 ms frames retained for the acoustic and inharmonicity analyses.}
    \label{fig:hnr_peak_selection_example}
\end{figure}

\subsection*{Vowel-Centred Frame Extraction}
\label{subsec:vowel_frame_extraction}

For the continuous speech, we initially extracted the voiced segments in order to allow for the vowel variations; these are then treated as independent vowel-centred analysis units.
 
These units are then used to form the subsequent acoustic and inharmonicity representations. To improve robustness, these representations should be computed on locally stable voiced material; the extraction procedure aims to determine such suitable target regions within the continuous utterances.

In order to determine which vowel that is spoken, the approximate word-level timing was obtained using the Vosk program\footnote{Vosk is an offline open-source speech recognition toolkit that provides word-level timestamps and confidence scores.\cite{vosk} 
}, using the appropriate language settings. 
The recogniser output was used as a coarse temporal guide. To ensure reliable vowel determination, words with low recognition confidence were discarded. The remaining orthographic forms were normalised to language-specific vowel categories, with accented or language-specific vowel characters mapped to corresponding canonical vowel labels. Each retained word interval then provided a restricted search region for the signal-based selection of vowel centres.

Within these word intervals, candidate centres were selected using their local harmonics-to-noise ratio (HNR), indicating regions with a clear periodic structure. 

The peak in the local HNR was deemed to indicate the candidate centres. A word was retained when the number of accepted centres match the number of vowels in its normalised orthographic form, and the centres were assigned vowel labels in their order of occurrence within the word. This produced an order-preserving vowel-centred approximation based on both the recognised word form and the local acoustic evidence.
Following conventional short-time assumption that speech is approximately stationary over windows of roughly 20--30\,ms,\cite{backstrom2022itsp} each 
accepted centre defined a 20\,ms frame.

Since HNR estimates depend on the analysis window and its duration,\cite{fernandes2018hnr} the local HNR search was performed before extracting the final 20\,ms frame. Frames with weak energy, excessive near-zero samples, or insufficient local support were removed. 
It is worth noting that such frame removal may omit relevant information, stressing the importance of the used weighting scheme. 
Overall, only a handful of the processed recordings did not retain any vowel-centred frames; as each user had made multiple recordings, no users were excluded as a result of this step.
%
 
The retained frames formed the common temporal support for the two continuous-speech representations.
Figure~\ref{fig:hnr_peak_selection_example} illustrates the HNR-based centre selection within a word segment.

\begin{figure}[t]
    \centering
    \includegraphics[width=\linewidth]{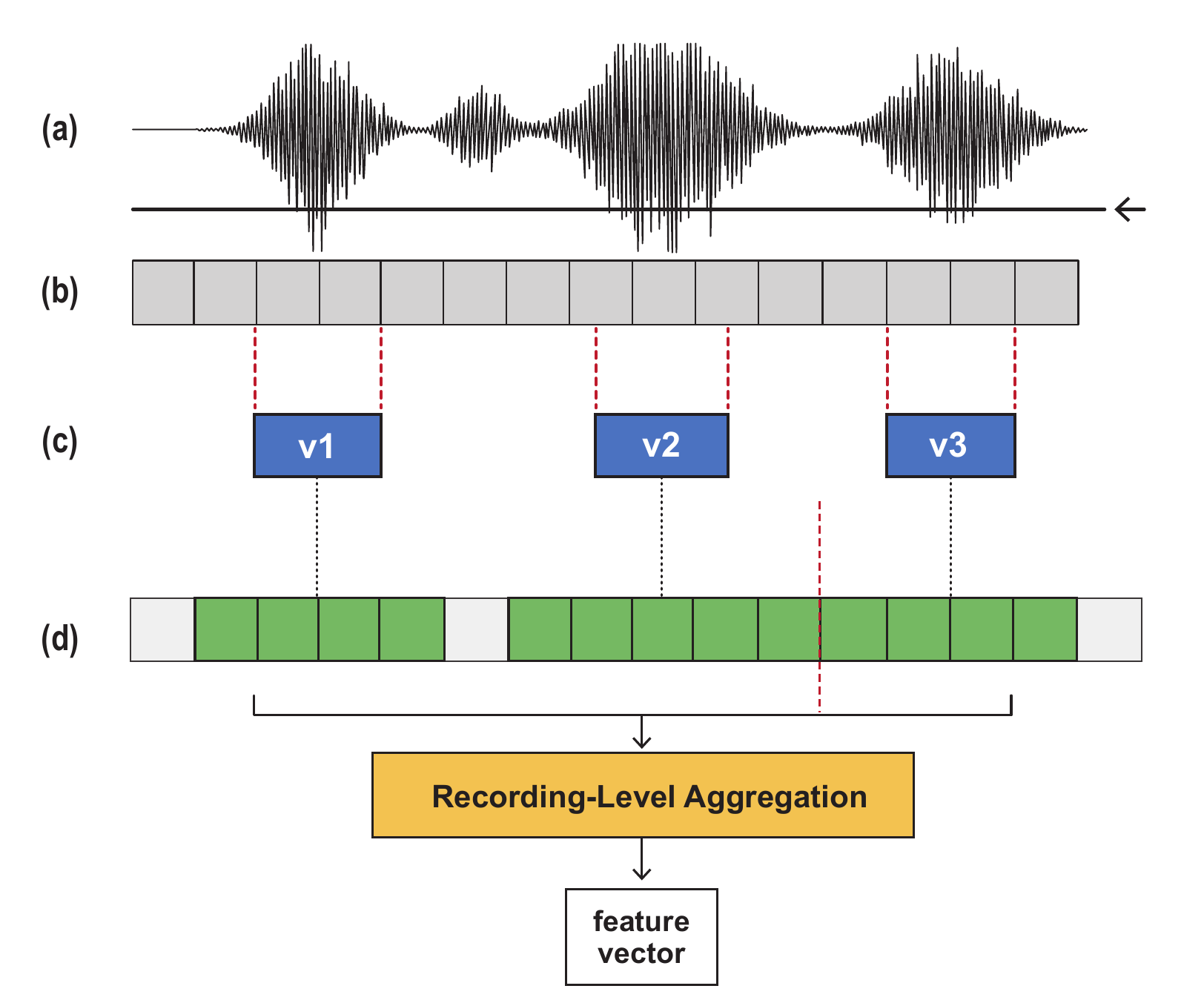}
    \caption{Schematic illustration of the continuous-speech acoustic representation. The full recording (a) was first analysed on the regular
    openSMILE short-time grid, producing an unaggregated stream of acoustic descriptor windows (b). Independently selected 20\,ms vowel-centred target frames (c) were then used as temporal masks: descriptor windows whose time intervals overlapped these target frames were retained (d), while non-overlapping windows were discarded. The retained descriptor windows were pooled within the recording and summarised by statistical functionals to form the final recording-level feature vector.}
    \label{fig:alignment_scheme}
\end{figure}

\subsection*{Acoustic Representation of Continuous Speech}
\label{subsec:continuous_acoustic_representation}

Using openSMILE\cite{eyben2010opensmile,eyben2016gemaps},
the 25 low-level descriptor (LLD) streams defined in the eGeMAPSv02 feature set were extracted from each full continuous-speech recording. These descriptors include loudness, fundamental frequency, jitter, shimmer, harmonics-to-noise ratio (HNR), alpha ratio, Hammarberg index, spectral slopes and spectral flux, MFCCs 1--4, harmonic spectral measures (H1--H2 and H1--A3), and
the frequencies, bandwidths, and relative amplitudes of the first three formants. 
The selected 20\,ms vowel-centred frames were then used as temporal masks to retain only the descriptor windows overlapping the selected material.

This kept the descriptor extraction on the regular openSMILE analysis grid while restricting the summary to the vowel-centred material selected in the previous step.
It should be noted that the selected frames and the openSMILE descriptor windows will generally use different time grids. To allow for this, these grids are linked by their temporal overlap. Let $I_i$ denote the $i$th selected target frame be centred at time $c_i$, with interval
\begin{equation}
    I_i = [c_i - 10,\, c_i + 10]\text{ ms}
    \label{eq:target_frame_interval}
\end{equation}
and $W_j$ the interval of the $j$th openSMILE analysis window in the same recording, with the overlap 
\begin{equation}
    \operatorname{ov}(i,j) = | I_i \cap W_j |.
    \label{eq:opensmile_overlap}
\end{equation}
A descriptor window was then retained if it overlapped at least one selected target frame; when a descriptor window overlapped multiple target frames, it was assigned to the target frame with the largest overlap,
\begin{equation}
    i^\ast(j) = \arg\max_i \operatorname{ov}(i,j).
    \label{eq:largest_overlap_assignment}
\end{equation}
Each retained descriptor window is only used once in the recording-level aggregation.
For each retained low-level descriptor, summary statistics were computed across the matched windows, including the mean, standard deviation,
median, 10th percentile, 90th percentile, and interquartile range. Because these overall summaries pool the retained vowel
material within a recording, they may also reflect differences in the vowel composition of that recording. To retain vowel-specific
information, the mean and standard deviation were additionally computed separately for each of the five vowel classes when sufficient selected frames were available. Additional support variables described the amount and distribution of the selected material, including the number of retained frames, the number of matched descriptor windows, temporal gap summaries, and local signal-quality summaries. Concatenating these quantities produced one fixed-length acoustic feature vector for each retained recording.
Together, these features describe each vowel's acoustic characteristics and the number of frames retained for it.
An overview of the process is shown in Figure~\ref{fig:alignment_scheme}.

\subsection*{Frame-Level Inharmonicity Estimation}
\label{subsec:inharmonicity_estimation}

The inharmonicity of a tonal signal is defined as the displacement of observed spectral peaks from an ideal harmonic grid.\cite{ChristensenJ09} Thus, for a harmonic signal with fundamental frequency \(f_0\), having its \(n\)th harmonic at \(n f_0\), the deviation of the $n$th estimated spectral peak, $\hat f_n$, from \(n f_0\) constitutes the $n$th inharmonicity offset, i.e., 
\begin{align}
    \delta_n = \hat f_n -  n f_0
\end{align}
It should be noted that the resulting offset estimate is strongly affected by any pitch errors, making this estimate critical for a reliable offset estimate. The use of short speech frames makes this especially problematic, since it is not uncommon that higher harmonics may dominate for such signals, 
which, together with limited frequency resolution, increases the risk of octave halving or doubling errors.\cite{backstrom2022itsp,elvander2020f0}

To limit this problem, we here form a two stage estimate of the inharmonicity offsets. For each retained vowel-centred frame, an autocorrelation estimate is used to provide a coarse initial pitch estimate. For a frame \(x[n]\), \(n=0,\ldots,N-1\), where \(N\) denotes the number of samples in the retained 20\,ms frame, the unscaled autocorrelation was computed as
\begin{equation}
    R_x[\tau] =
    \sum_{n=\tau}^{N-1} x[n]x[n-\tau],
    \label{eq:autocorrelation_pitch}
\end{equation}
from which the dominant lag within a plausible speech pitch range of 50--500\,Hz was used to form an initial pitch estimate. The corresponding lag bounds were
\begin{equation}
    \tau_{\min}=\left\lfloor \frac{f_s}{500} \right\rfloor,
    \qquad
    \tau_{\max}=\min\left(\left\lfloor \frac{f_s}{50} \right\rfloor,N-1\right).
\end{equation}
Thus,
\begin{equation}
    \hat\tau_{0}
    =
    \arg\max_{\tau \in [\tau_{\min},\tau_{\max}]}
    R_x[\tau],
    \qquad
    \hat f_{0}^{init} = \frac{f_s}{\hat\tau_{0}}.
    \label{eq:autocorrelation_mode}
\end{equation}
This estimate is then used to restrict the subsequent search interval
\begin{table}[t]
\centering
\caption{Inharmonicity data on recording- and person-level.}
\label{tab:inh_representation_overview}
\footnotesize
\begin{tabular}{lcc}
\toprule
NeuroVoz data set & Recording level & Person level \\
\midrule
Number of analysis units & 1683 & 107 \\
Mean total usable frames & 8.26 & 129.85 \\
Median total usable frames & 8 & 133 \\
Mean core-complete frames & 5.00 & 78.67 \\
Median core-complete frames & 5 & 80 \\
\toprule
Voice Diagnostics data set & Recording level& Person level \\
\midrule
Number of analysis units & 3560 & 175 \\
Mean total usable frames & 61.99 & 1261.13 \\
Median total usable frames & 61 & 430 \\
Mean core-complete frames & 31.73 & 645.51 \\
Median core-complete frames & 33 & 128 \\
\bottomrule
\end{tabular}
\end{table}
\begin{equation}
    f_0 \in
    [f_{\min},f_{\max}]
    \cap
    [(1-\rho)\hat f_{0}^{init} ,(1+\rho)\hat f_{0}^{init} ],
    \label{eq:local_pitch_search_band}
\end{equation}
where \(f_{\min}=50\)\,Hz, \(f_{\max}=500\)\,Hz, and \(\rho=0.25\). If no reliable autocorrelation estimate was obtained, a fallback interval of 70--350\,Hz was used. This restricted search was used to reduce octave-halving and octave-doubling errors while keeping the estimate within a physiologically plausible speech-pitch range. Within the resulting interval, the pitch estimate was initialised by a least-squares harmonic-grid search over 200 uniformly spaced candidate frequencies using the first six harmonics, and was then refined using a regularised harmonic-grid fit.

In this work, we formed this refinement using the near-harmonic optimal mass transport (OMT) estimator introduced previously,\cite{elvander2020f0,elvander2023otp} in which
the sinusoidal components are expected to lie close to a harmonic grid, 
such that
\begin{equation}
    x[n]
    \approx
    \sum_{k=1}^{K}
    a_k
    \cos\!\left(
    2\pi \frac{k f_0 + \delta_k}{f_s} n + \phi_k
    \right),
    \label{eq:regularised_harmonic_model}
\end{equation}
where \(a_k\) and \(\phi_k\) denote the amplitude and phase of the \(k\)th component, and \(\delta_k\) a (possible) small deviation from the exact location of the harmonic \(k f_0\). Using the approach of Elvander,\cite{elvander2023otp} the sought parameters are determined as the solution to
\begin{equation}
    \min_{f_0,\{\delta_k\},\theta}
    \left\|
    x - \hat{x}(f_0,\delta,\theta)
    \right\|_2^2
    +
    \lambda
    \sum_{k=1}^{K} w_k \delta_k^2 ,
    \label{eq:regularised_grid_fit}
\end{equation}
where \(\theta\) collects the remaining sinusoidal parameters and \(\lambda\) controls the regularisation strength and \(w_k\) denotes the amplitude-dependent penalty weight. The amplitude weighting corresponds to the local quadratic form of the OMT prior for small inharmonic perturbations, favouring a near-harmonic grid while still allowing small deviations where needed. 
The resulting estimates are then combined with the earlier autocorrelation-based priors to allow for a subsequent octave-correction step to improve robustness for short frames.

This is done by comparing the estimates with octave-shifted alternatives and with the autocorrelation-based initial estimate, \(\hat f_{0}^{init}\), forming the candidate set
\begin{equation}
    \mathcal{C}
    =
    \left\{
    \hat f_{\mathrm{AH}},\;
    0.5\hat f_{\mathrm{AH}},\;
    2\hat f_{\mathrm{AH}},\;
    \hat f_{0}^{init},\;
    0.5\hat f_{0}^{init},\;
    2\hat f_{0}^{init}
    \right\}
    \label{eq:octave_candidate_set}
\end{equation}
Following Klapuri,\cite{klapuri2003multiple} 
each candidate was scored by the accumulated spectral evidence around its first \(K\) nominal
harmonic locations,
\begin{equation}
    S(f)
    =
    \sum_{k=1}^{K}
    \frac{1}{k}
    \max_{\nu \in \mathcal{B}_k(f)}
    |X(\nu)|,
    \label{eq:multiharmonic_score}
\end{equation}
where \(X(\nu)\) denotes the short-time magnitude spectrum
at frequency \(\nu\). 
For a candidate fundamental frequency
\(f\), the nominal location of the \(k\)th harmonic is \(kf\).
The local search band was defined as
\begin{table}[t]
\centering
\caption{Performance of the five sustained-vowel benchmark models from the NeuroVoz data set. 
}
\label{tab:sustained_vowel_results}
\footnotesize
\begin{tabular}{lcccc}
\toprule
Vowel & Person AUC & Person F1 & Rec. AUC & Rec. F1 \\
\midrule
$[a]$ & 0.58 $\pm$ 0.05 & 0.60 $\pm$ 0.07 & 0.58 $\pm$ 0.03 & 0.66 $\pm$ 0.05 \\
$[e]$ & 0.63 $\pm$ 0.09 & 0.68 $\pm$ 0.02 & 0.67 $\pm$ 0.04 & 0.73 $\pm$ 0.06 \\
$[i]$ & 0.77 $\pm$ 0.11 & 0.70 $\pm$ 0.07 & 0.77 $\pm$ 0.09 & 0.74 $\pm$ 0.07 \\
$[o]$ & 0.77 $\pm$ 0.07 & 0.69 $\pm$ 0.10 & 0.73 $\pm$ 0.07 & 0.73 $\pm$ 0.08 \\
$[u]$ & \textbf{0.83 $\pm$ 0.12} & 0.64 $\pm$ 0.07 & \textbf{0.80 $\pm$ 0.12} & 0.62 $\pm$ 0.04 \\
\bottomrule
\end{tabular}
\end{table}
\begin{equation}
    \mathcal{B}_k(f)
    =
    \left\{
    \nu \in [0,f_s/2]:
    |\nu-kf| \leq b_k(f)
    \right\},
    \label{eq:harmonic_search_band}
\end{equation}
with the empirical half-width
\begin{equation}
    b_k(f)=\max\{12\,\mathrm{Hz},\,0.18f,\,0.06kf\}.
\end{equation}
This choice provided a minimum absolute tolerance for
short-frame spectral resolution, a relative tolerance around
the candidate fundamental, and a gradually wider absolute
tolerance for higher-order harmonics.
%
%
The fundamental frequency was then selected as
\begin{equation}
    \widehat{f}_0
    =
    \arg\max_{f \in \mathcal{C}} S(f).
    \label{eq:final_pitch_selection}
\end{equation}
after which the sought inharmonicity coefficients were determined from the spectral offset from $k \widehat{f}_0$, with the $n$th harmonic frequency being determined as
%
\begin{equation}
    \hat f_n
    =
    \arg\max_{f \in [n\widehat{f}_0-b_n,\;n\widehat{f}_0+b_n]}
    |X(f)|,
    \label{eq:local_harmonic_peak}
\end{equation}
with $b_n$ allowing for an estimate in the search-band in \eqref{eq:harmonic_search_band}, yielding 

\begin{equation}
    \hat\delta_n = f_n - n\widehat{f}_0
    \label{eq:signed_offset}
\end{equation}
In this work, only the first $P=6$ harmonic orders were retained, since higher orders were deemed less stable for the considered short 
frames, yielding the inharmonicity features
\begin{equation}
    \boldsymbol{\delta}_t =
    \begin{bmatrix}
    \hat\delta_{1,t} & \ldots 
    & \hat\delta_{P,t}
    \end{bmatrix}^{\top}.
    \label{eq:offset_vector}
\end{equation}
Only frames passing quality-control criteria were used in the statistical representations introduced below. Retained frames required root-mean-square (RMS) energy, computed as \(20\log_{10}(\sqrt{N^{-1}\sum_n x[n]^2}+\epsilon)\),
of at least \(-85\)\,dB, a maximum normalised autocorrelation peak of at least 0.60 within the 50--500\,Hz pitch-lag range, and a finite pitch estimate \(\widehat f_0\). Frames were discarded if the final pitch estimate was octave-inconsistent with the autocorrelation-based initial estimate \((\widehat f_0 < 0.70\widehat f_0^{init}\) or
\(\widehat f_0 > 1.60\widehat f_0^{init})\). All the $P$ retained harmonic offsets also had to be finite and lie within their local search bands, \(|\widehat{\delta}_k| \leq b_k(\widehat f_0)\).

\begin{table}[t]
\centering
\caption{Performance of the sustained-vowel benchmark models from the Voice Diagnostic data set. }
\label{tab:sustained_vowel_results_VD}
\footnotesize
\begin{tabular}{lcccc}
\toprule
Vowel & Person AUC & Person F1 & Rec. AUC & Rec. F1 \\
\midrule
$[a]$ & 0.77 $\pm$ 0.04 & 0.49 $\pm$ 0.05 & 0.64 $\pm$ 0.14 & 0.38 $\pm$ 0.13 \\
\bottomrule
\end{tabular}
\end{table}

\subsection*{Person-Level Inharmonicity Representation}
\label{subsec:person_level_inharmonicity}

Given the restrictive selection of the inharmonicity frames, a single short recording may only yield a few reliable estimates. In order to allow for reliable estimates of the statistical properties of these frames, the inharmonicity representations from all retained recordings from the same speaker are pooled.

For speaker $s$, let $\mathcal{R}_s$ denote the set of retained 
recordings, and let $\mathcal{D}_r$ denote the set of estimated inharmonicity frames
from recording $r$. The speaker-level pool was defined as
\begin{equation}
    \mathcal{D}_s
    =
    \bigcup_{r \in \mathcal{R}_s} \mathcal{D}_r
    =
    \left\{
    \boldsymbol{\delta}^{(s)}_t
    \right\}_{t=1}^{M_s},
    \qquad
    \boldsymbol{\delta}^{(s)}_t \in \mathbb{R}^{P}.
    \label{eq:person_level_pool}
\end{equation}
where $M_s$ denotes the total number of inharmonicity frames for speaker $s$.
Furthermore, define the corresponding speaker-level mean offset vector
\begin{equation}
    \boldsymbol{\mu}_s
    =
    \frac{1}{M_s}
    \sum_{t=1}^{M_s}
    \boldsymbol{\delta}^{(s)}_t .
    \label{eq:person_mean_offset}
\end{equation}
The centred offset vectors were then used to describe the spread and shape of the speaker's offset cloud. 

Let ${\bf R}_s$ denote the sample covariance matrix for speaker $s$. Then, the (robust) $P \times P$ Ledoit--Wolf shrinkage  covariance matrix estimate, $\widetilde{\boldsymbol{\Sigma}}_s$, is formed as
\begin{equation}
    \widetilde{\boldsymbol{\Sigma}}_s
    =
    (1-\gamma_s) {\bf R}_s
    +
    \gamma_s
    \frac{\operatorname{tr}({\bf R}_s)}{P} {\bf I}_P ,
    \label{eq:ledoit_wolf_person}
\end{equation}
where \({\bf I}_P\) denotes the \(P\times P\) identity matrix and the shrinkage intensity \(0 \leq \gamma_s \leq 1\) was estimated from the speaker's retained offset vectors. 
The resulting estimate preserves the average variance of \({\bf R}_s\) while shrinking the covariance structure toward an isotropic form.
Table~\ref{tab:inh_representation_overview} summarizes the retained data on both recording- and person-level.

In summary, the estimation thus proceeds in three main steps. First, a stable fundamental frequency is estimated for each retained vowel-centred frame. Second, the estimated spectral components are compared with their nominal harmonic locations to obtain signed offsets from the ideal harmonic grid. Finally, these frame-level offsets are pooled across the retained recordings of
each speaker and summarised through their distributional and geometric structure, providing the person-level inharmonicity representation used for classification.

\section{Benchmark, Classification, and Fusion}
\label{sec:classification_fusion}

Proceeding, the above representations are converted into speaker-level PD probabilities using the evaluation protocol introduced in Section~\ref{sec:data_protocol}. The sustained-vowel benchmark and the continuous-speech acoustic analysis were both trained from recording-level acoustic features which were then aggregated for each speaker. The inharmonicity analysis was fitted directly at the person level, since its input representation had already been pooled across the retained recordings of each speaker.
The speaker-level probability was obtained by averaging over that speaker's recording-level probabilities, corresponding to the post-mean aggregation strategy described previously,\cite{yang2025aggregation} i.e., 

\begin{equation}
    \widehat{p}_s
    =
    \frac{1}{|\mathcal{R}_s|}
    \sum_{r \in \mathcal{R}_s}
    \widehat{p}_r .
    \label{eq:recording_to_person_aggregation}
\end{equation}

For all reported binary classification measures, the operating threshold was selected within each evaluation fold using the validation speakers only. The threshold, $\tau^*$, was chosen to maximise the speaker-level validation \(F_1\) score, i.e., 
\begin{equation}
    \tau^\ast
    =
    \arg\max_{\tau \in \mathcal{T}_{\mathrm{val}}}
    F_{1,\mathrm{val}}(\tau),
    \label{eq:validation_threshold_selection}
\end{equation}
where \(\mathcal{T}_{\mathrm{val}}\) denotes the candidate thresholds evaluated on the validation speakers, and
\begin{equation}
F_{1,\mathrm{val}}(\tau)
=
\frac{2\,TP}
{2\,TP+FP+FN} ,
\end{equation}
with the true positives (\(TP\)), false positives (\(FP\)), and false negatives (\(FN\)) being computed at the speaker level on the validation set using the threshold $\tau$. The selected threshold was then applied unchanged to the held-out test speakers in the corresponding evaluation fold.

\begin{table}[t]
\centering
\caption{Previous NeuroVoz sustained-vowel results.\cite{ozbolt2022methodological}}
\label{tab:ozbolt_neurovoz_sustained_results}
\footnotesize
\begin{tabular}{lcc}
\toprule
Vowel & Best accuracy  & Best setting \\
\midrule
$[a]$ & 75.6 & DARTH-VAT + FS-RFC \\
$[e]$ & 74.5 & DARTH-VAT + RFC \\
$[i]$ & 81.8 & DARTH-VAT + FS-RFC \\
$[o]$ & 76.3 & DARTH-VAT + FS-RFC \\
$[u]$ & 83.6 & DARTH-VAT + RFC \\
\bottomrule
\end{tabular}
\end{table}

\begin{table}[t]
\centering
\caption{Previous reported VD results.\cite{momeni2026reliable}}
\label{tab:vd_external_momeni}
\footnotesize
\begin{tabular}{lcccc}
\toprule
Model & Acc. & F1 & Recall & Precision \\
\midrule
DistilHuBERT & 70.2 & 72.5 & 70.2 & 76.2 \\
XGBoost      & 59.3 & 63.8 & 59.3 & 78.7 \\
TabNet       & 66.4 & 70.0 & 66.4 & 79.8 \\
\bottomrule
\end{tabular}
\end{table}

\subsection*{The Sustained-Vowel Benchmark}
\label{subsec:sustained_benchmark}

In order to evaluate the improvement of using continuous speech in place of a sustained-vowel sound, we create a benchmark using the best performing vowel sound. 
As each sustained-vowel recording contains one production of a single vowel, no additional frame selection or alignment was needed for these recordings, and the eGeMAPSv02 features were computed directly from each recording using openSMILE,
after the fold-local preprocessing as described in Section~\ref{subsec:evaluation_protocol}.
For the NeuroVoz data set, separate models were trained for the five sustained vowels $[a]$, $[e]$, $[i]$, $[o]$, and $[u]$; as the VD data set only contains the $[a]$ vowel, a single model was trained for this data set.

Proceeding, recording-level classifiers were formed using  
XGBoost,\cite{chen2016xgboost} which provides a flexible model for tabular acoustic features and can accommodate nonlinear feature effects and interactions. As a result, a tree-boosting model was used as the acoustic benchmark while keeping the classifier family aligned with the continuous-speech acoustic model. For a recording-level feature vector, \(x_i\), the model output was formed as
\begin{equation}
    \widehat{p}_i
    =
    \sigma\!\left(
    \sum_{m=1}^{M} f_m(x_i)
    \right),
    \label{eq:xgboost_probability}
\end{equation}
where \(f_m(\cdot)\) denotes the contribution of the \(m\)th boosted tree and \(\sigma(\cdot)\) is the logistic function. Model tuning and early stopping were carried out within the validation procedure. Class imbalance was handled through class-dependent training weights, and final performance was reported after the recording-level probabilities had been formed using 
\eqref{eq:recording_to_person_aggregation}.

\begin{figure}[t]
    \centering
    \includegraphics[width=\linewidth]{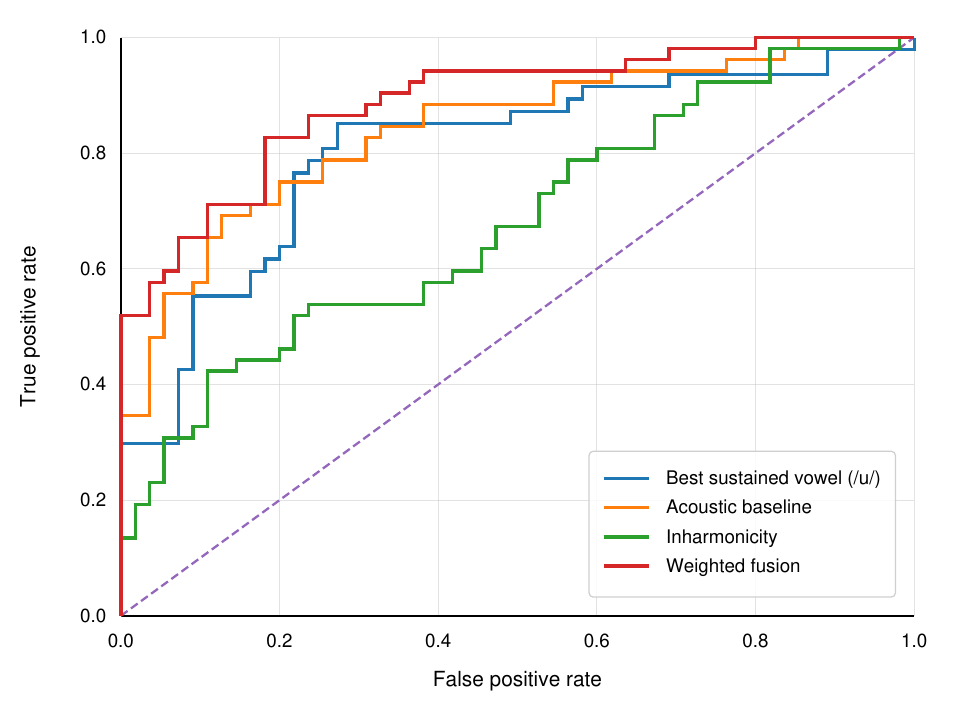}
    \caption{Pooled out-of-fold person-level ROC comparison for the NeuroVoz data, showing the sustained-vowel benchmark ($[u]$) as compared to the continuous-speech acoustic model, the inharmonicity model, and the weighted fusion model.}
    \label{fig:main_model_roc}
\end{figure}

\subsection*{Continuous-Speech Acoustic Model}
\label{subsec:continuous_acoustic_model}

The continuous-speech acoustic model used the recording-level features extracted from the vowel-centred regions, as detailed  in Section~\ref{subsec:continuous_acoustic_representation}. The modelling choices were kept as close to the sustained-vowel benchmark as possible to allow for a fair comparision on the speech representations rather than on a change of classifier structure. Thus, both classifiers were formed using XGBoost on the acoustic tabular features, and both produced recording-level probabilities before speaker-level aggregation. 
%
As the continuous speech data are less uniform than the sustained-vowel sounds, with speakers contributing varying numbers of recordings, the diagnostic classes were not  balanced. To allow for this, the training weights combined class balancing with the speaker-balancing principle, as described in Section~\ref{subsec:evaluation_protocol}, to ensure that a speaker with many recordings did not disproportionally affect the classification. 

The XGBoost model was fitted to the recording-level training data within each outer fold. The operating-threshold selection and early stopping used the speaker-level validation split; the recording-level validation probabilities were first averaged for each speaker, with the validation performance being computed on the resulting person-level probabilities. The same aggregation was then applied to the held-out test speakers, i.e.,
\begin{equation}
    p^{\mathrm{acoustic}}_s
    =
    \frac{1}{|\mathcal{R}^{\mathrm{LR}}_s|}
    \sum_{r \in \mathcal{R}^{\mathrm{LR}}_s}
    \widehat{p}^{\mathrm{acoustic}}_r ,
    \label{eq:continuous_acoustic_person_probability}
\end{equation}
where \(\mathcal{R}^{\mathrm{LR}}_s\) is the set of retained recordings for speaker \(s\). The threshold selection and the final evaluation were then performed at the person level.

\begin{figure}[t]
    \centering
    \includegraphics[width=\linewidth]{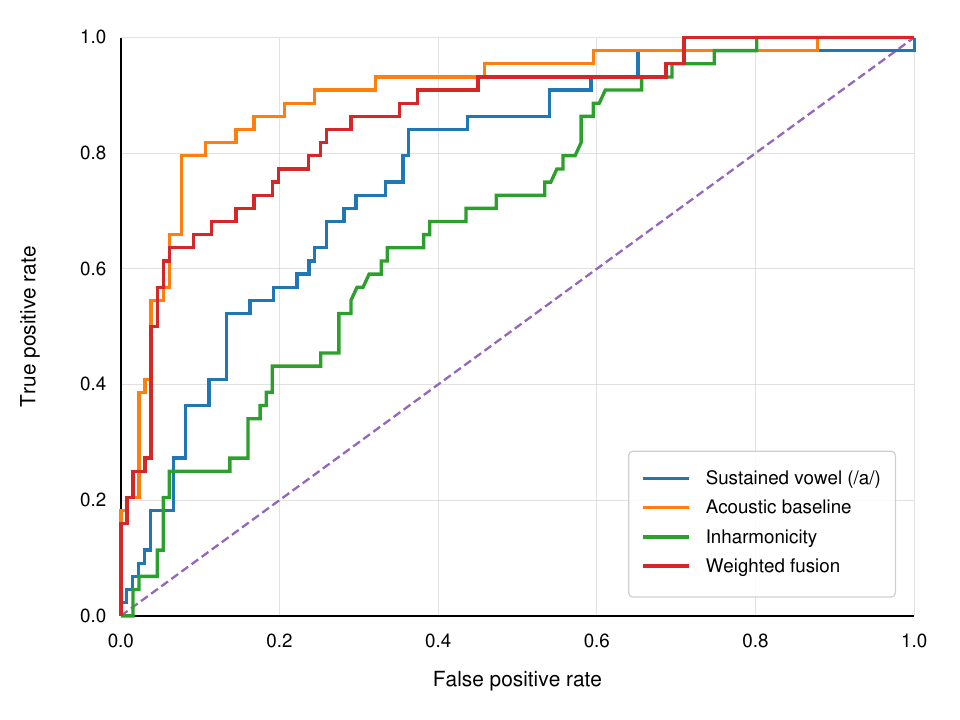}
    \caption{Pooled out-of-fold person-level ROC comparison for the VD data, showing the sustained-vowel benchmark ($[a]$) as compared to the continuous-speech acoustic model, the inharmonicity model, and the weighted fusion model.}
    \label{fig:main_model_roc_VD}
\end{figure}

\begin{table*}[t]
\centering
\caption{Person-level performance of the continuous-speech models. Values are reported as mean $\pm$ standard deviation across the five outer folds.}
\label{tab:continuous_speech_results}
\footnotesize
\begin{tabular}{lccccc}
\toprule
Model (NeuroVoz) & AUC & F1 & ACC & Recall & Specificity \\
\midrule
Acoustic model & 0.86 $\pm$ 0.05 & 0.73 $\pm$ 0.03 & 0.70 $\pm$ 0.07 & 0.81 $\pm$ 0.13 & 0.60 $\pm$ 0.24 \\
Inharmonicity model & 0.73 $\pm$ 0.01 & 0.65 $\pm$ 0.06 & 0.61 $\pm$ 0.07 & 0.75 $\pm$ 0.15 & 0.47 $\pm$ 0.19 \\
Weighted fusion & \textbf{0.90 $\pm$ 0.06} & \textbf{0.75 $\pm$ 0.06} & 0.74 $\pm$ 0.05 & \textbf{0.85 $\pm$ 0.19} & 0.64 $\pm$ 0.21 \\
Logistic fusion & 0.82 $\pm$ 0.04 & 0.71 $\pm$ 0.06 & \textbf{0.74 $\pm$ 0.04} & 0.70 $\pm$ 0.15 & \textbf{0.78 $\pm$ 0.07} \\
\toprule
Model (VD) & AUC & F1 & ACC & Recall & Specificity \\
\midrule
Acoustic model & \textbf{0.91 $\pm$ 0.04} & 0.72 $\pm$ 0.04 & \textbf{0.85 $\pm$ 0.04} & 0.80 $\pm$ 0.15 & 0.86 $\pm$ 0.10 \\
Inharmonicity model & 0.65 $\pm$ 0.14 & 0.38 $\pm$ 0.13 & 0.56 $\pm$ 0.12 & 0.58 $\pm$ 0.32 & 0.55 $\pm$ 0.22 \\
Weighted fusion & 0.85 $\pm$ 0.09 & 0.70 $\pm$ 0.09 & \textbf{0.85 $\pm$ 0.03} & 0.75 $\pm$ 0.16 & \textbf{0.88 $\pm$ 0.03} \\
Logistic fusion & 0.90 $\pm$ 0.05 & \textbf{0.74 $\pm$ 0.05} & \textbf{0.85 $\pm$ 0.04} & \textbf{0.87 $\pm$ 0.08} & 0.84 $\pm$ 0.06 \\
\bottomrule
\end{tabular}
\end{table*}

\subsection*{Person-Level Inharmonicity Model}
\label{subsec:inharmonicity_model}

The inharmonicity representation enters the classification at the person level. 
Using the covariance matrix estimate 
$\widetilde{\boldsymbol{\Sigma}}_s$ obtained from  \eqref{eq:ledoit_wolf_person}, several potential inharmonicity features were formed. In order to determine suitable candidate features from this set, we performed a feature reduction using sparse logistic regression together with cross-validation, retaining only features deemed relevant for all folds. The resulting features are summarized in the Appendix. 

As the retained features are correlated, these were then combined using an elastic-net logistic-regression model, providing some shrinkage while retaining sparsity.\cite{zou2005elasticnet,friedman2010glmnet}

For speaker \(i\), let \(z_i \in \mathbb{R}^{q}\) denote the condensed feature vector, where \(q=|\mathcal{F}_{\mathrm{stable}}|\). The raw person-level probabilities,
\begin{equation}
    p^{\mathrm{raw}}_i
    =
    \sigma(\beta_0 + z_i^\top \boldsymbol{\beta})
    =
    \frac{1}{
    1+\exp[-(\beta_0 + z_i^\top \boldsymbol{\beta})]
    } ,
    \label{eq:elastic_net_probability}
\end{equation}
were then used to determine the model parameters by minimizing
\begin{equation}
\begin{split}
    \min_{\beta_0,\boldsymbol{\beta}}
    \Bigg[
    &-
    \sum_{i \in \mathcal{T}}
    w_i
    \left\{
    y_i \log p^{\mathrm{raw}}_i
    +
    (1-y_i)\log(1-p^{\mathrm{raw}}_i)
    \right\} \\
    &+
    \lambda
    \left(
    \rho \|\boldsymbol{\beta}\|_1
    +
    \frac{1-\rho}{2}
    \|\boldsymbol{\beta}\|_2^2
    \right)
    \Bigg],
\end{split}
\label{eq:elastic_net_objective}
\end{equation}
where \(\mathcal{T}\) denotes the set of training speakers, \(w_i\) the class weights, whereas \(\lambda\) controls the overall penalty strength and \(\rho \in [0,1]\) the balance between the \(L_1\) and \(L_2\) parts of the penalty. The hyperparameters were selected on the validation speakers using validation AUC.

The amount of frame-level data supporting the person-level differed across speakers, implying that speakers with more frames provided a more stable estimate of the covariance than speakers with a more limited frame support. To reflect this, the raw inharmonicity probability was adjusted after classification with  $n_i$, the number of included frames for speaker $i$, such that the
adjusted probability was formed as
\begin{equation}
    p_i^{\mathrm{inharm}}
    =
    0.5
    +
    \frac{n_i}{n_i+k}
    \left(
    p_i^{\mathrm{raw}} - 0.5
    \right)
    \label{eq:inharmonicity_reliability_adjustment}
\end{equation}
where the parameter $k$ controls the strength of the shrinkage; this was selected on the validation speakers using the validation AUC after adjustment.
%
%
This reliability adjustment follows the general shrinkage principle that estimates supported by less information should be pulled toward a more conservative value.\cite{datta2012smallarea} The adjusted probability \(p_i^{\mathrm{inharm}}\) was used for threshold selection, test evaluation, and fusion.

\subsection*{Score-Level Fusion}
\label{subsec:score_level_fusion}

As the acoustic and inharmonicity models describe different aspects of the speech signal, they will partly complement each other. The acoustic model captures a broader set of short-time descriptors from the selected vowel-centred regions, whereas the inharmonicity model focuses on the person-level inharmonicity structure. To allow for both aspects, the two models are merged in the final stage. Score-level fusion is a transparent way to combine classifiers that rely on different representations while producing comparable decision scores.\cite{kittler1998combining}
The fused person-level probability was computed as a weighted average,
\begin{equation}
    p^{\mathrm{fused}}_s
    =
    w\,p^{\mathrm{acoustic}}_s
    +
    (1-w)\,p^{\mathrm{inharm}}_s,
    \label{eq:weighted_fusion}
\end{equation}
with the relative weighting $w \in [0,1]$ being selected on the validation speakers within each outer fold. After the fused validation probabilities had been obtained, the final decision threshold was also selected on the validation speakers and then applied unchanged to the held-out test speakers.

\section{Results}
\label{sec:results}

In order to evaluate the gain of using the continuous speech models, we initially examine the performance of the sustained-vowel benchmarks for each of the two data sets. Proceeding, we then evaluate the continuous-speech acoustic and inharmonicity representations, before examining whether their fusion improves
person-level PD classification in the two data sets.
Performance was reported primarily at the person level using the area under the receiver operating characteristic curve (AUC) and F1-score. The AUC was used as a threshold-free ranking measure, whereas F1 reflected the binary speaker-level decision after validation-based threshold selection.

\subsection*{Sustained-Vowel Benchmark Results}
\label{subsec:sustained_vowel_results}

Tables~\ref{tab:sustained_vowel_results} and \ref{tab:sustained_vowel_results_VD} show the mean and standard deviation across the five outer folds for the single-vowel sustained-phonation models for the NeuroVoz and VD data sets, respectively. As seen in Table~\ref{tab:sustained_vowel_results} the different vowels in the NeuroVoz data set show distinctly different performance, with $[u]$ offering the most reliable classification results, with a mean person-level AUC of 0.83.
The $[o]$ and $[i]$ models formed the next tier, whereas $[a]$ and $[e]$ show weaker performance for this dataset\footnote{These results were also confirmed using pairwise permutation tests, with $[u]$ exceeding $[a]$ by 0.236 AUC points (\(p=0.0012\)) and $[e]$ by 0.185 AUC points (\(p=0.0048\)). The corresponding differences relative to $[i]$ and $[o]$ were smaller and not statistically supported by the tests (\(p=0.2737\) and \(p=0.3791\), respectively).}, which is especially noteworthy as $[a]$ is the most common vowel sound used  in speech-based PD studies. 
A similar vowel-wise pattern was also reported by Ozbolt {\em et al.}~\cite{ozbolt2022methodological} for the same data set, as summarized in Table~\ref{tab:ozbolt_neurovoz_sustained_results}. As the reported results from that study\cite{ozbolt2022methodological} are at the recording level, not speaker level as here, and use somewhat different controls for class, age, and gender effects from our formulation, it is difficult to fully compare the results, although it is clear from their work that $[u]$ was also seen to yield the most reliable representation, confirming our results.

Compared with the results for $[a]$ for the VD data set, the sustained $[a]$ benchmark performed better than the NeuroVoz $[a]$ model, but given that the data sets differ in language, cohort, and recording setting, one should again be careful to draw any conclusions from this comparison.
It may be noted that the results reported here are in line with the record-level results presented for this data set by Momeni {\em et al.}\cite{momeni2026reliable}, as summarized in Table~\ref{tab:vd_external_momeni}; these results are also formed somewhat differently than the here presented results, making direct comparison difficult. 
To allow for the best possible performance in the following comparison, we use $[u]$ as the vowel benchmark model for the  NeuroVoz data set; as only $[a]$ is available for the VD set, this is the vowel benchmark for this data\footnote{It may be noted that two multi-vowel sustained-phonation variants were also examined for the NeuroVoz data. A pooled-vowel model reached a person-level AUC of 0.78 and F1 of 0.70. An equal-weight probability fusion across the five vowel-specific models reached a person-level AUC of 0.83 and F1 of 0.75. These results show that combining sustained vowels can improve the sustained-vowel reference, particularly in F1.}.

\subsection*{Continuous-Speech Model Results}
\label{subsec:continuous_speech_results}

Table~\ref{tab:continuous_speech_results} reports the person-level results for the main continuous-speech models in both data sets. In NeuroVoz, the continuous-speech acoustic model achieved a mean AUC of 0.857 and F1 of 0.726, exceeding the best single-vowel sustained benchmark before any inharmonicity information was added. The inharmonicity model was weaker as a standalone classifier, with a mean AUC of 0.725 and F1 of 0.647.
The similar pattern was observed in the VD data set. The continuous-speech acoustic model reached a mean AUC of 0.908 and F1 of 0.724, clearly above the sustained $[a]$ benchmark reported in Table~\ref{tab:sustained_vowel_results_VD}; the standalone inharmonicity model again performed substantially below the acoustic model.
Fusing the continuous speech model with the inharmonicity information gave differing results for the two data sets; for NeuroVoz, the weighted score fusion gave the highest observed AUC and F1, reaching a mean AUC of 0.899 and F1 of 0.752. 
However, for the VD data set, the acoustic-only model gave the highest AUC, and the inclusion of the inharmonicity information did not offer a similar improvement as was seen from the NeuroVoz data set, suggesting that the improvement from the inharmonicity information is more data-dependent, perhaps due to the difference between the listen-repeat and the continuous speech utterances.

Figures~\ref{fig:main_model_roc} and~\ref{fig:main_model_roc_VD} show the pooled out-of-fold ROC curves for the main models, with the NeuroVoz results showing the noted improvement from the weighted fusion over the acoustic model, whereas the VD curves show that the acoustic model already captured most of the discriminative information available in that data set. 
Although it is clear that the continuous speech model clearly outperforms the sustained vowel approach, further studies on the value of incorporating the inharmonicity information are clearly necessary.

It should be noted that the obtained results are comparable to the listen-repeat results reported for the continuous speech model presented by Postma and Tejedor-Garcia\cite{postma2025evaluating}, as summarized in Table~\ref{tab:postma_neurovoz_lr_results}. The best model in this study uses pre-trained full-recording audio embeddings employed on 1~s frames with 0.1~s overlap, computing 6144 features for each, which are then combined with an SVM classifier\footnote{In contrast, it may be noted that our model only use 116 and 140 recording-level features for the NeuroVoz and VD data sets, respectively.}. 
Although the presented results are similar between this and our work, one should be careful in comparing these, because Postma and Tejedor-Garcia\cite{postma2025evaluating} evaluated predictions at the recording level, whereas we aggregated recording-level probabilities and computed the primary
performance measures at the speaker level. Furthermore, the work by Postma and Tejedor-Garcia\cite{postma2025evaluating} does not account for the notable differences in age between the HC and PD users, nor take into account the different number of recordings between groups (see also Table~\ref{tab:dataset_summary}; this
perhaps explains the low agreement between models and demographic difficulties noted by Postma and Tejedor-Garcia\cite{postma2025evaluating}); further studies should thus be made to allow for a fair comparison between the methods.
However, one notable difference between the methods presented by Postma and Tejedor-Garcia\cite{postma2025evaluating} and the here proposed vowel-centred acoustic and inharmonicity framework worth noting is that the latter offers the benefit of allowing for a direct acoustic interpretation, which is difficult to form from the full-recording embeddings.

\section{Discussion}
\label{sec:discussion_conclusion}

\subsection*{Interpretation of the Main Findings}
\label{subsec:discussion_main_findings}

The clearest finding in this study was the advantage of using the continuous-speech acoustic representation over the sustained-vowel reference. The continuous-speech acoustic model notably exceeded the benchmark under the same speaker-level protocol, confirming the previously reported results.\cite{postma2025evaluating} This agreement across two different datasets suggests that PD classification models should preferably be formed using continuous speech. 

In this work, the recordings were not treated as uniform acoustic objects; rather, to allow for improved interpretability, they were first reduced to vowel-centred regions with sufficient local support for short-time analysis. This step is introduced as the utterances also contains consonants, pauses, transitions, and weakly voiced regions. The performance of the acoustic model, therefore, reflects both the use of continuous speech and the attempt to focus its representation on locally stable voiced material.
The inharmonicity results require a more cautious interpretation. The inharmonicity model did not match the continuous-speech acoustic model as a standalone classifier for either data set. However, the weighted score fusion clearly improved the results for the 
NeuroVoz data set, although no such gains could be seen for the VD data set. However,  the inclusion did not significantly weaken the results for the VD data set either,  suggesting that it may be possible to include the inharmonicity information for another data set without risking any decremental results. However, 
further studies are required to trust such conclusions and on the gain of incorporating the inharmonicity information in the model.

\begin{table}[t]
\centering
\caption{Previous NeuroVoz continuous speech results for the listen-repeat recordings.\cite{postma2025evaluating}}
\label{tab:postma_neurovoz_lr_results}
\footnotesize
\begin{tabular}{llcc}
\toprule
Embedding & Classifier & ACC & AUC \\
\midrule
OpenL3 & SVM & \textbf{0.82 $\pm$ 0.04} & \textbf{0.90 $\pm$ 0.03} \\
OpenL3 & KNN & 0.76 $\pm$ 0.03 & 0.81 $\pm$ 0.03 \\
OpenL3 & ERT & 0.71 $\pm$ 0.02 & 0.80 $\pm$ 0.02 \\
VGGish & SVM & 0.78 $\pm$ 0.05 & 0.85 $\pm$ 0.05 \\
VGGish & KNN & 0.72 $\pm$ 0.03 & 0.77 $\pm$ 0.04 \\
VGGish & ERT & 0.73 $\pm$ 0.03 & 0.79 $\pm$ 0.04 \\
Wav2Vec2.0 & SVM & 0.76 $\pm$ 0.03 & 0.84 $\pm$ 0.02 \\
Wav2Vec2.0 & KNN & 0.74 $\pm$ 0.02 & 0.78 $\pm$ 0.03 \\
Wav2Vec2.0 & ERT & 0.70 $\pm$ 0.02 & 0.77 $\pm$ 0.03 \\
\bottomrule
\end{tabular}
\end{table}

\subsection*{Methodological Implications}
\label{subsec:discussion_methodological_implications}

It is important to stress that the comparison between sustained vowels and continuous speech was made under a common evaluation protocol. This was necessary because most speakers contributed multiple recordings, and because preprocessing choices can otherwise introduce information from held-out speakers. The speaker-level split, fold-local preprocessing, validation-based threshold selection, and person-level reporting were therefore part of the experimental design rather than secondary implementation details. Without this structure, the comparison between speech tasks would be difficult to interpret.
The results also support the use of local analysis units in continuous speech. Whole-recording summaries are simple to form, but would combine acoustically different regions of the utterance. The vowel-centred representation used here gave a more targeted way to exploit continuous speech while retaining compatibility with established acoustic descriptors.
The inharmonicity analysis points to a different modelling issue: the level at which the representation should be formed. Individual continuous speech recordings provided too few complete offset vectors for stable covariance- and shape-based summaries. Pooling across recordings at the person level gave more support for these descriptors and matched the final evaluation target. Even so, the weaker standalone performance of the inharmonicity model indicates that harmonic-offset features are more sensitive to frame availability, pitch-grid stability, and dataset structure than the broader acoustic representation.

\subsection*{Limitations and Future Work}
\label{subsec:discussion_limitations_future}

As all studies, our study has clear limitations; notably, the two examined data sets differ in language and speech tasks. NeuroVoz used Castilian-Spanish \emph{listen-repeat} utterances, whereas the VD data set used spontaneous continuous speech in Swedish. This difference is helpful to examine the robustness of the drawn conclusions, although it also limits precise task-level comparisons. Future work with harmonised sustained-vowel and continuous-speech protocols across languages would help separate language effects, task effects, and representation effects more cleanly.

Furthermore, some parts of the continuous-speech framework involved empirical design choices. This applies in particular to the vowel-frame selection procedure and to the construction of the geometry-based inharmonicity feature set. These choices were appropriate for the aim and settings of the present study, but should not be treated as uniquely determined solutions. More robust frame selection, improved handling of ASR uncertainty, and alternative aggregation strategies may further improve the continuous-speech representation.
Recording conditions are another limitation. NeuroVoz had a reported mean  signal-to-noise ratio (SNR) of 24.3~dB \cite{mendes2024neurovozpaper}, while the VD recordings were collected using the participants' own smartphones in various settings. Background noise and device variability may have affected acoustic feature estimation. We also note that, for instance, HNR estimates obtained using the unmodified openSMILE implementation may also be less reliable in speakers with severe dysphonia or under noisy recording conditions, potentially affecting vowel-frame selection and the resulting feature summaries.
Finally, the available  metadata also limits the clinical interpretation of the present findings.


\section{Conclusion}
\label{subsec:discussion_conclusion_final}

This study examines whether a continuous speech model may offer preferable performance for voiced-based Parkinson's disease classification as compared to traditional sustained-vowel models. Using two distinctly different data sets, we clearly show that a carefully designed continuous-speech model outperforms the best sustained-vowel benchmarks.
We show how continuous speech may be efficiently represented to provide added predictive value when evaluated under a strict speaker-level framework. We further introduce an inharmonicity information model that shows promise as a complementary representation, improving the classification performance, although further studies are required to allow for firm conclusions.

\appendices

\section{Retained Inharmonicity Feature Sets}
\label{app:feature_sets}
\setcounter{table}{0}
\renewcommand{\thetable}{A-\Roman{table}}

The person-level inharmonicity models used sparse, data set specific
subsets of the same candidate descriptor families. Table~\ref{tab:appendix_inh_selected_features}
lists the retained descriptors. Here, corr \(i\)--\(j\) denotes the
correlation between harmonic-offset orders \(i\) and \(j\).

\begin{table}[t]
\centering
%
\scriptsize
\refstepcounter{table}
\label{tab:appendix_inh_selected_features}
\textbf{TABLE~\thetable}\\[-0.2em]
\textsc{Retained Person-Level Inharmonicity Descriptors After Sparse Feature Selection}
\vspace{0.35em}

\begin{tabular}{@{}p{0.15\columnwidth}p{0.48\columnwidth}p{0.28\columnwidth}@{}}
\toprule
Data set & Retained descriptor(s) & Interpretation \\
\midrule
& Cross-vowel SD of mean absolute correlation
& Cross-vowel dependence variability \\

& Median centred-offset radius; quartile skewness of centred-offset radius
& Radial spread and asymmetry \\

& Cross-vowel SD of mean-offset norm; signed balance of the mean-offset vector
& Offset-location variability and direction \\

NeuroVoz
& Corr \(1\)--\(2\), \(1\)--\(5\), \(1\)--\(6\), \(2\)--\(3\), \(2\)--\(4\), \(2\)--\(5\), and \(3\)--\(4\)
& Pairwise harmonic dependence \\

& \(/a/\)-conditioned mean absolute correlation
& Vowel-conditioned dependence strength \\

& \(/o/\)-conditioned leading eigenvalue ratio
& Vowel-conditioned covariance shape \\

& Maximum absolute lag-3 correlation; mean lag-4 correlation; mean and median lag-5 correlation
& Lagged correlation summaries \\

\midrule

& Median raw offset-vector norm
& Typical uncentred offset size \\

& \(/e/\)- and \(/o/\)-conditioned mean absolute correlation
& Vowel-conditioned dependence strength \\

& \(/e/\)-conditioned covariance log determinant; \(/e/\)-conditioned leading eigenvalue ratio
& Vowel-conditioned covariance size and shape \\

Voice Diagnostics
& \(/u/\)-conditioned 90th percentile radius
& Vowel-conditioned high radial spread \\

& Covariance flatness
& Offset-cloud isotropy \\

& Maximum absolute lag-1 correlation; mean absolute lag-4 correlation; mean and median lag-5 correlation
& Lagged correlation summaries \\

& Corr \(1\)--\(6\)
& Pairwise harmonic dependence \\
\bottomrule
\end{tabular}
\end{table}

\section*{Data Availability} The NeuroVoz corpus is publicly available as described in the cited dataset publication. The Voice Diagnostics dataset is not publicly available due to privacy agreements and institutional data-sharing restrictions.

\bibliographystyle{vancouver}
\bibliography{reference}

@article{cao2025speech,
  author  = {Cao, F. and Vogel, A. P. and Gharahkhani, P. and Renteria, M. E.},
  title   = {Speech and language biomarkers for {Parkinson's} disease prediction, early diagnosis and progression},
  journal = {npj Parkinson's Disease},
  volume  = {11},
  pages   = {57},
  year    = {2025},
  doi     = {10.1038/s41531-025-00913-4}
}

@article{hlavnicka2017connected,
  author  = {Hlavni{\v{c}}ka, J. and {\v{C}}mejla, R. and Tykalov{\'{a}}, T. and {\v{S}}onka, K. and R{\r{u}}{\v{z}}i{\v{c}}ka, E. and Rusz, J.},
  title   = {Automated analysis of connected speech reveals early biomarkers of {Parkinson's} disease in patients with rapid eye movement sleep behaviour disorder},
  journal = {Scientific Reports},
  volume  = {7},
  pages   = {12},
  year    = {2017},
  doi     = {10.1038/s41598-017-00047-5}
}

@article{vaiciukynas2017speech,
  author  = {Vaiciukynas, E. and Verikas, A. and Gelzinis, A. and Bacauskiene, M.},
  title   = {Detecting {Parkinson's} disease from sustained phonation and speech signals},
  journal = {PLOS ONE},
  volume  = {12},
  number  = {10},
  pages   = {e0185613},
  year    = {2017},
  doi     = {10.1371/journal.pone.0185613}
}

@article{momeni2025gws,
  author  = {Momeni, N. and Whitling, S. and Jakobsson, A.},
  title   = {Interpretable {Parkinson's} Disease Detection Using Group-Wise Scaling},
  journal = {IEEE Access},
  volume  = {13},
  pages   = {29147--29161},
  year    = {2025},
  doi     = {10.1109/ACCESS.2025.3540600}
}

@article{appakaya2023connected,
  author  = {Appakaya, S. B. and Pratihar, R. and Sankar, R.},
  title   = {{Parkinson's} Disease Classification Framework Using Vocal Dynamics in Connected Speech},
  journal = {Algorithms},
  volume  = {16},
  number  = {11},
  pages   = {509},
  year    = {2023},
  doi     = {10.3390/a16110509}
}

@article{ozbolt2022methodological,
  author  = {Ozbolt, A. S. and Moro-Vel{\'{a}}zquez, L. and Lina, I. and Butala, A. A. and Dehak, N.},
  title   = {Things to Consider When Automatically Detecting {Parkinson's} Disease Using the Phonation of Sustained Vowels: Analysis of Methodological Issues},
  journal = {Applied Sciences},
  volume  = {12},
  number  = {3},
  pages   = {991},
  year    = {2022},
  doi     = {10.3390/app12030991}
}

@article{ge2023evaluation,
  author  = {Ge, W. and Lueck, C. and Suominen, H. and Apthorp, D.},
  title   = {Has machine learning over-promised in healthcare?: A critical analysis and a proposal for improved evaluation, with evidence from {Parkinson's} disease},
  journal = {Artificial Intelligence in Medicine},
  volume  = {139},
  pages   = {102524},
  year    = {2023},
  doi     = {10.1016/j.artmed.2023.102524}
}

@article{mendes2024neurovozpaper,
  author  = {Mendes-Laureano, J. and G{\'{o}}mez-Garc{\'{i}}a, J. A. and Guerrero-L{\'{o}}pez, A. and Luque-Buzo, E. and Arias-Londo{\~{n}}o, J. D. and Grandas-P{\'{e}}rez, F. J. and Godino-Llorente, J. I.},
  title   = {{NeuroVoz}: a {Castillian} {Spanish} corpus of parkinsonian speech},
  journal = {Scientific Data},
  volume  = {11},
  pages   = {1367},
  year    = {2024},
  doi     = {10.1038/s41597-024-04186-z}
}

@article{brabenec2017speech,
  author  = {Brabenec, L. and Mekyska, J. and Galaz, Z. and Rektorova, I.},
  title   = {Speech disorders in {Parkinson's} disease: early diagnostics and effects of medication and brain stimulation},
  journal = {Journal of Neural Transmission},
  volume  = {124},
  number  = {3},
  pages   = {303--334},
  year    = {2017},
  doi     = {10.1007/s00702-017-1676-0}
}

@article{atalar2023hypokinetic,
  author  = {Atalar, M. S. and Oguz, O. and Genc, G.},
  title   = {Hypokinetic Dysarthria in {Parkinson's} Disease: A Narrative Review},
  journal = {Sisli Etfal Hastanesi Tip Bulteni},
  volume  = {57},
  number  = {2},
  pages   = {163--170},
  year    = {2023},
  doi     = {10.14744/SEMB.2023.29560}
}

@article{eyben2016gemaps,
  author  = {Eyben, F. and Scherer, K. R. and Schuller, B. W. and Sundberg, J. and Andr{\'{e}}, E. and Busso, C. and Devillers, L. Y. and Epps, J. and Laukka, P. and Narayanan, S. S. and Truong, K. P.},
  title   = {The {Geneva Minimalistic Acoustic Parameter Set} ({GeMAPS}) for Voice Research and Affective Computing},
  journal = {IEEE Transactions on Affective Computing},
  volume  = {7},
  number  = {2},
  pages   = {190--202},
  year    = {2016},
  doi     = {10.1109/TAFFC.2015.2457417}
}

@inproceedings{eyben2010opensmile,
  author    = {Eyben, F. and W{\"{o}}llmer, M. and Schuller, B.},
  title     = {{openSMILE}: The {Munich} Versatile and Fast Open-Source Audio Feature Extractor},
  booktitle = {Proceedings of the 18th ACM International Conference on Multimedia},
  pages     = {1459--1462},
  year      = {2010},
  doi       = {10.1145/1873951.1874246}
}

@article{elvander2020f0,
  author  = {Elvander, F. and Jakobsson, A.},
  title   = {Defining Fundamental Frequency for Almost Harmonic Signals},
  journal = {IEEE Transactions on Signal Processing},
  volume  = {68},
  pages   = {6453--6466},
  year    = {2020},
  doi     = {10.1109/TSP.2020.3035466}
}

@inproceedings{elvander2023otp,
  author    = {Elvander, F.},
  title     = {Estimating Inharmonic Signals with Optimal Transport Priors},
  booktitle = {2023 IEEE International Conference on Acoustics, Speech and Signal Processing},
  pages     = {1--5},
  year      = {2023},
  doi       = {10.1109/ICASSP49357.2023.10095082}
}

@inproceedings{chen2016xgboost,
  author    = {Chen, T. and Guestrin, C.},
  title     = {{XGBoost}: A Scalable Tree Boosting System},
  booktitle = {Proceedings of the 22nd ACM SIGKDD International Conference on Knowledge Discovery and Data Mining},
  pages     = {785--794},
  year      = {2016},
  doi       = {10.1145/2939672.2939785}
}

@article{yang2025aggregation,
  author  = {Yang, Z. and Zhou, H. and Srivastav, S. and Shaffer, J. G. and Abraham, K. E. and Naandam, S. M. and Kakraba, S.},
  title   = {Optimizing {Parkinson's} Disease Prediction: A Comparative Analysis of Data Aggregation Methods Using Multiple Voice Recordings via an Automated Artificial Intelligence Pipeline},
  journal = {Data},
  volume  = {10},
  number  = {1},
  pages   = {4},
  year    = {2025},
  doi     = {10.3390/data10010004}
}

@book{backstrom2022itsp,
  author    = {B{\"{a}}ckstr{\"{o}}m, T. and R{\"{a}}s{\"{a}}nen, O. and Zewoudie, A. and P{\'{e}}rez Zarazaga, P. and Koivusalo, L. and Das, S. and G{\'{o}}mez Mellado, E. and Bouafif Mansali, M. and Ramos, D.},
  title     = {Introduction to Speech Processing: 2nd Edition},
  publisher = {Zenodo},
  year      = {2022},
  doi       = {10.5281/zenodo.6821775}
}

@article{fernandes2018hnr,
  author  = {Fernandes, J. and Teixeira, F. and Guedes, V. and Junior, A. and Teixeira, J. P.},
  title   = {Harmonic to Noise Ratio Measurement -- Selection of Window and Length},
  journal = {Procedia Computer Science},
  volume  = {138},
  pages   = {280--285},
  year    = {2018},
  doi     = {10.1016/j.procs.2018.10.040}
}

@article{friedman2010glmnet,
  author  = {Friedman, J. and Hastie, T. and Tibshirani, R.},
  title   = {Regularization Paths for Generalized Linear Models via Coordinate Descent},
  journal = {Journal of Statistical Software},
  volume  = {33},
  number  = {1},
  pages   = {1--22},
  year    = {2010},
  doi     = {10.18637/jss.v033.i01}
}

@article{zou2005elasticnet,
  author  = {Zou, H. and Hastie, T.},
  title   = {Regularization and Variable Selection via the Elastic Net},
  journal = {Journal of the Royal Statistical Society: Series B},
  volume  = {67},
  number  = {2},
  pages   = {301--320},
  year    = {2005},
  doi     = {10.1111/j.1467-9868.2005.00503.x}
}

@article{datta2012smallarea,
  author  = {Datta, G. and Ghosh, M.},
  title   = {Small Area Shrinkage Estimation},
  journal = {Statistical Science},
  volume  = {27},
  number  = {1},
  pages   = {95--114},
  year    = {2012},
  doi     = {10.1214/11-STS374}
}

@article{kittler1998combining,
  author  = {Kittler, J. and Hatef, M. and Duin, R. P. W. and Matas, J.},
  title   = {On Combining Classifiers},
  journal = {IEEE Transactions on Pattern Analysis and Machine Intelligence},
  volume  = {20},
  number  = {3},
  pages   = {226--239},
  year    = {1998},
  doi     = {10.1109/34.667881}
}

@misc{parkinsonFoundationStats,
  author = {{Parkinson's Foundation}},
  title  = {{Statistics}},
  year   = {2024},
  url    = {https://www.parkinson.org/understanding-parkinsons/statistics},
  note   = {Accessed: 5 February 2024}
}

@article{kalia2015parkinsons,
  author  = {Kalia, L. V. and Lang, A. E.},
  title   = {{Parkinson's disease}},
  journal = {The Lancet},
  volume  = {386},
  number  = {9996},
  pages   = {896--912},
  year    = {2015},
  doi     = {10.1016/S0140-6736(14)61393-3}
}

@article{mei2021machine,
  author  = {Mei, J. and Desrosiers, C. and Frasnelli, J.},
  title   = {{Machine learning for the diagnosis of Parkinson's disease: A review of literature}},
  journal = {Frontiers in Aging Neuroscience},
  volume  = {13},
  pages   = {633752},
  year    = {2021},
  doi     = {10.3389/fnagi.2021.633752}
}

@article{moro2021advances,
  author  = {Moro-Vel{\'a}zquez, L. and G{\'o}mez-Garc{\'i}a, J. A. and Arias-Londo{\~n}o, J. D. and Dehak, N. and Godino-Llorente, J. I.},
  title   = {{Advances in Parkinson's disease detection and assessment using voice and speech: A review of the articulatory and phonatory aspects}},
  journal = {Biomedical Signal Processing and Control},
  volume  = {66},
  pages   = {102418},
  year    = {2021},
  doi     = {10.1016/j.bspc.2021.102418}
}

@incollection{gullapalli2022early,
  author    = {Gullapalli, A. S. and Mittal, V. K.},
  title     = {{Early detection of Parkinson's disease through speech features and machine learning: A review}},
  booktitle = {{ICT with Intelligent Applications}},
  series    = {{Smart Innovation, Systems and Technologies}},
  volume    = {248},
  publisher = {Springer},
  address   = {Singapore},
  year      = {2022},
  doi       = {10.1007/978-981-16-4177-0_22}
}

@inproceedings{momeni2024mobile,
  author    = {Momeni, Niloofar and Whitling, Susanna and Jakobsson, Andreas},
  title     = {{Detecting Parkinson's Disease Using Voice Recordings From Mobile Devices}},
  booktitle = {Proceedings of the 32nd European Signal Processing Conference},
  pages     = {1516--1520},
  year      = {2024},
  doi       = {10.23919/EUSIPCO63174.2024.10715471}
}

@misc{who2023parkinson,
  author = {{World Health Organization}},
  title  = {{Parkinson's Disease}},
  year   = {2023},
  url    = {https://www.who.int/news-room/fact-sheets/detail/parkinson-disease},
  note   = {Accessed: 5 February 2024}
}

@article{murman2012early,
  author  = {Murman, Daniel L.},
  title   = {{Early treatment of Parkinson's disease: Opportunities for managed care}},
  journal = {The American Journal of Managed Care},
  volume  = {18},
  number  = {7 Suppl},
  pages   = {S183--S188},
  year    = {2012}
}

@article{schrag2000quality,
  author  = {Schrag, Anette and Jahanshahi, Marjan and Quinn, Niall},
  title   = {{How does Parkinson's disease affect quality of life? A comparison with quality of life in the general population}},
  journal = {Movement Disorders},
  volume  = {15},
  number  = {6},
  pages   = {1112--1118},
  year    = {2000},
  month   = nov,
  doi     = {10.1002/1531-8257(200011)15:6<1112::AID-MDS1008>3.0.CO;2-A}
}

@inproceedings{vasquezcorrea2015continuous,
  author    = {V{\'a}squez-Correa, Juan Camilo and Arias-Vergara, Tom{\'a}s and Orozco-Arroyave, Juan Rafael and Vargas-Bonilla, Jes{\'u}s Francisco and Arias-Londo{\~n}o, Juli{\'a}n David and N{\"o}th, Elmar},
  title     = {{Automatic Detection of Parkinson's Disease from Continuous Speech Recorded in Non-Controlled Noise Conditions}},
  booktitle = {{Proceedings of Interspeech 2015}},
  pages     = {105--109},
  year      = {2015},
  doi       = {10.21437/Interspeech.2015-36}
}

@article{orozcoarroyave2016running,
  author  = {Orozco-Arroyave, Juan Rafael and H{\"o}nig, Florian and Arias-Londo{\~n}o, Juli{\'a}n David and Vargas-Bonilla, Jes{\'u}s Francisco and Daqrouq, Khalid and Skodda, Sabine and Rusz, Jan and N{\"o}th, Elmar},
  title   = {{Automatic Detection of Parkinson's Disease in Running Speech Spoken in Three Different Languages}},
  journal = {The Journal of the Acoustical Society of America},
  volume  = {139},
  number  = {1},
  pages   = {481--500},
  year    = {2016},
  doi     = {10.1121/1.4939739}
}

@article{farago2023cnn,
  author  = {Farag{\'o}, Paul and {\c{S}}tef{\u{a}}nig{\u{a}}, Sebastian-Aurelian and Cordo{\c{s}}, Claudia-Georgiana and Mih{\u{a}}il{\u{a}}, Laura-Ioana and Hintea, Sorin and Pe{\c{s}}tean, Ana-Sorina and Beyer, Michel and Perju-Dumbrav{\u{a}}, L{\u{a}}cr{\u{a}}mioara and Ile{\c{s}}an, Robert Radu},
  title   = {{CNN-Based Identification of Parkinson's Disease from Continuous Speech in Noisy Environments}},
  journal = {Bioengineering},
  volume  = {10},
  number  = {5},
  pages   = {531},
  year    = {2023},
  doi     = {10.3390/bioengineering10050531}
}

@article{idrisoglu2026multiclass,
  author  = {Idrisoglu, Alper and Behrens, Anders},
  title   = {{Use of machine learning and voice for multiclass classification of Parkinson's disease, chronic obstructive pulmonary disease, and healthy controls}},
  journal = {Scientific Reports},
  volume  = {16},
  pages   = {15485},
  year    = {2026},
  doi     = {10.1038/s41598-026-53409-3}
}

@misc{vosk,
  author = {{Alpha Cephei}},
  title  = {{Vosk Speech Recognition Toolkit}},
  year   = {2024},
  url    = {https://alphacephei.com/vosk/},
  note   = {Accessed: 24 May 2026}
}

@article{klapuri2003multiple,
  author  = {Klapuri, Anssi P.},
  title   = {{Multiple Fundamental Frequency Estimation Based on Harmonicity and Spectral Smoothness}},
  journal = {IEEE Transactions on Speech and Audio Processing},
  volume  = {11},
  number  = {6},
  pages   = {804--816},
  year    = {2003},
  month   = nov,
  doi     = {10.1109/TSA.2003.815516}
}

@book{ChristensenJ09,
  author    = {Christensen, Mads Gr{\ae}sb{\o}ll and Jakobsson, Andreas},
  title     = {Multi-Pitch Estimation},
  series    = {Synthesis Lectures on Speech and Audio Processing},
  volume    = {5},
  publisher = {Morgan \& Claypool Publishers},
  address   = {San Rafael, CA},
  year      = {2009},
  doi       = {10.2200/S00178ED1V01Y200903SAP005},
  isbn      = {9781598298383}
}

@inproceedings{postma2025evaluating,
author    = {Postma, Emmy and Tejedor-Garcia, Cristian},
title     = {Evaluating the Effectiveness of Pre-Trained Audio Embeddings for Classification of {P}arkinson's Disease Speech Data},
booktitle = {Proceedings of Interspeech},
year      = {2025},
pages     = {4603--4607},
doi       = {10.21437/Interspeech.2025-801},
issn      = {2958-1796}
}

@inproceedings{momeni2026reliable,
  author    = {Momeni, Niloofar and Whitling, Susanna and Jakobsson, Andreas},
  title     = {Reliable AI via Age-Balanced Validation: Fair Model Selection for Parkinson's Detection from Voice},
  booktitle = {ICASSP 2026 -- 2026 IEEE International Conference on Acoustics, Speech and Signal Processing (ICASSP)},
  year      = {2026},
  pages     = {3031--3035},
  doi       = {10.1109/ICASSP55912.2026.11464625},
  isbn      = {979-8-3315-6701-9},
  publisher = {IEEE}
}

@article{rusz2013vowel,
  author  = {Rusz, Jan and Cmejla, Roman and Tykalova, Tereza and Ruzickova, Hana and Klempir, Jiri and Majerova, Veronika and Picmausova, Jana and Roth, Jan and Ruzicka, Evzen},
  title   = {Imprecise Vowel Articulation as a Potential Early Marker of {Parkinson's} Disease: Effect of Speaking Task},
  journal = {The Journal of the Acoustical Society of America},
  volume  = {134},
  number  = {3},
  pages   = {2171--2181},
  year    = {2013},
  doi     = {10.1121/1.4816541}
}

@article{illner2023vowel,
  author  = {Illner, Vojtech and Tykalova, Tereza and Skrabal, Dominik and Klempir, Jiri and Rusz, Jan},
  title   = {Automated Vowel Articulation Analysis in Connected Speech Among Progressive Neurological Diseases, Dysarthria Types, and Dysarthria Severities},
  journal = {Journal of Speech, Language, and Hearing Research},
  volume  = {66},
  number  = {8},
  pages   = {2600--2621},
  year    = {2023},
  doi     = {10.1044/2023_JSLHR-22-00526}
}

@article{farajollahi2026diagnosis,
  author  = {Farajollahi, Boshra and Saffarian, Arezoo and Habibi, Seyed Amir Hassan and Sheikhtaheri, Abbas},
  title   = {Diagnosis of Parkinson's Disease Based on Voice and Speech Using Machine Learning and Deep Learning: A Systematic Umbrella Review},
  journal = {Journal of Voice},
  year    = {2026},
  doi     = {10.1016/j.jvoice.2026.06.055},
  note    = {Online ahead of print}
}

@article{pedregosa2011scikit,
  title={Scikit-learn: Machine Learning in Python},
  author={Pedregosa, F. and Varoquaux, G. and Gramfort, A. and others},
  journal={Journal of Machine Learning Research},
  volume={12},
  pages={2825--2830},
  year={2011}
}

\end{document}